\documentclass[12pt,preprint]{aastex}

\newcommand{\dgr}{$^{\circ}$}
\newcommand{\etal}{et~al.\ }

\usepackage[twocolumn]{emulateapj5}

\slugcomment{Scheduled to appear in The Astrophysical Journal (v569, April 10, 2002)}

\shorttitle{Polarization of 2MASS QSOs}
\shortauthors{Smith \etal}

\begin{document}

\title{The Optical Polarization of Near-Infrared Selected QSO{\scriptsize{S}}}

\author{Paul S. Smith, Gary D. Schmidt, and Dean C. Hines}
\affil{Steward Observatory, The University of Arizona,
    Tucson, AZ 85721}

\and

\author{Roc M. Cutri and Brant O. Nelson}
\affil{California Institute of Technology, IPAC, MS 100--22,
Pasadena, CA 91125}

\begin{abstract}

Optical broad-band polarimetry is presented for near-infrared color-selected
active galactic nuclei (AGN)
classified as quasi-stellar objects (QSOs)
based on their $K_s\/$-band luminosity.
More than 10\% of a sample of 70 QSOs discovered in the Two Micron All Sky
Survey (2MASS) with $J - K_s > 2$ and $M_{K_s} < -23$
show high broad-band linear polarization ($P > 3$\%), and values range to 
a maximum of $P \sim 11$\%.
High polarization tends to be associated with the most luminous
objects at $K_s\/$, and with QSOs having the highest near-IR--to--optical
flux ratios. 
The 2MASS QSO sample includes objects possessing a wide range 
of optical spectral
types.
High polarization is seen in two of 22
broad emission-line (Type~1) objects, but
$\sim 1/4$ of the QSOs of intermediate spectral type (Type~1.5--1.9) are
highly polarized.
None of the nine QSOs classified as Type~2 exhibit $P > 3$\%.
It is likely that 
the unavoidable inclusion of unpolarized starlight from the host galaxy
within the observation aperture
results in reduced
polarization for the narrow emission-line objects.

The high polarization of 2MASS-discovered QSOs
supports the conclusion
inferred from their near-IR and optical colors, that the nuclei of 
many of these objects are obscured to some degree by dust.
Correlations
between optical polarization and near-IR luminosity and color imply
that the dominant polarizing mechanism in the sample is scattering
of AGN light into our line of sight
by material located in relatively unobscured regions near the nucleus.
The broad-band polarization properties of the 2MASS QSO sample are
compared to those of other, predominantly radio-quiet, QSOs and
are found to be consistent with the idea that the orientation of AGN
to the line of sight plays a major role in determining their
observed properties.

\end{abstract}


\section{Introduction}

Despite several decades of intense scrutiny, our 
understanding of active galactic nuclei (AGN) remains incomplete.
Fundamental issues such as the number density of AGN, their
contributions to the far-infrared and X-ray backgrounds, the role of
orientation to our line of sight, the processes that form and
power the nuclei, and their evolution still need to be resolved.
Recent large-scale X-ray, optical, and near-infrared surveys have
begun to reveal new populations of AGN
with number densities comparable to, or exceeding,
the number density of AGN derived from previous optical and radio
surveys \citep[e.g.,][]{sanders96}.
In particular, the Two Micron All Sky Survey
\citep[2MASS;][]{skrutskie97}
is finding large numbers
of AGN that have been missed by traditional UV/optical search
techniques.
New survey wavebands can uncover
past observational biases and reveal a wider range
of AGN properties.

Polarization is an important property that can be used to
explore the physics and structure of AGN, and
polarimetry yields valuable information about the various classes of
active nuclei.
For instance, early surveys of the most luminous known AGN --
quasi-stellar objects (QSOs) and
quasi-stellar radio sources (quasars) --
showed that they could be divided into two classes by their optical
polarization properties \citep{moore84}.
The vast majority of objects, both radio and optically selected, show
low optical polarization ($P < 3$\%; \citealt{stockman84}).
About 1\%
of the sample exhibited higher and variable polarization
\citep{moore81}, in part earning them the designation of optically
violent variable quasars (OVVs).
Almost all of these objects are associated with compact radio
sources, and their variability behavior, both in flux and in polarization,
clearly signifies that optically thin synchrotron radiation dominates the
UV/optical continua. 

Identification of the polarization mechanism(s) in
low-polarization QSOs has not been as straightforward largely
because of the low levels of polarization encountered.
Except for
a handful of broad absorption-line
QSOs (BALQSOs; \citealt{moore84,schmidt99}) and QSOs discovered by the
{\it Infrared Astronomical Satellite\/} ({\sl IRAS\/}) \citep{hines94},
only the OVVs exhibit polarization
above the traditional high-polarization threshold of $P = 3$\%.
A good illustration of the difficulty in identifying the polarizing
mechanism in most QSOs is provided by the
blue-excess Palomar-Green (PG) sample
\citep{schmidt83}.
\citet{berriman90} have shown that this predominantly radio-quiet sample
has a mean intrinsic optical polarization
of only 0.5\%
and the maximum observed polarization is 2.5\%.
Unambiguous statistical tests
designed to determine the source of the polarized flux
have been extremely hard to obtain,
even though high quality broad-band polarimetry
exists for the entire PG QSO sample.

This is not the case for the
the lower luminosity analogs to radio-quiet QSOs, the Seyfert~1
galaxies.
These objects show overall polarization properties similar to the
PG QSOs, but their polarization distribution extends to higher values.
The high-polarization Seyfert~1 ``tail'' better facilitates comparison
of the polarization distribution with dust
extinction and dust emission indicators,
and led \citet{berriman89} to
conclude that the primary polarizing mechanism in Seyfert~1s is
the scattering of nuclear light by dust located within the
narrow emission-line region (NLR).
Variations of this mechanism have been successful in
explaining the polarization of several Seyfert~2 galaxies and showing that
these objects would look like Seyfert~1s if viewed from
sightlines that intercept the scattering
material and the NLR
\citep[e.g.,][]{antonucci85,millgood90,tran95b,
wilkes95}.

{\sl IRAS\/} found a small
number of infrared-luminous AGN which suggest the
existence of a population of
red, optically-obscured radio-quiet QSOs
\citep{beichman86,low88,low89}.
Spectropolarimetry of several of these
``hyperluminous'' infrared galaxies (HIGs) has confirmed that their
tremendous IR luminosities are powered in large part
by QSOs hidden from our direct view,
and that polarization arises from scattering of nuclear light by
material with a relatively unobstructed view of the AGN
\citep{wills92,hines95a,hines95b,hines99,hines01,goodrich96,
young96,tran00}. 
A similar situation has been found for several radio
galaxies \citep{alighieri94,jannuzi95,cohen99}.
For these objects, light scattered by dust located up to several kiloparsecs
from the nucleus reveals the rest-frame ultraviolet spectrum of a QSO hidden
from our direct view.

\citet{cutri01} have initiated
a near-infrared, high galactic latitude survey for AGN using
2MASS.
The survey is designed to find new red, radio-quiet objects
and has, in fact, revealed a large population of
previously unidentified AGN.
The new sample generally consists of low-redshift ($z < 0.7$)
objects with redder near-IR colors
(a criterion of the object selection) and
much higher near-IR--to--optical flux ratios than optically
selected AGN.
Many of these AGN can be classified as QSOs in their own right since
their $K_s\/$-band luminosities are comparable to those of optically
selected QSOs.

Higher levels of optical polarization relative to
UV-excess QSOs are expected
from the 2MASS sample if, as with the IR-selected HIGs,
many of these objects are obscured along our line of sight.
Nuclear light scattered
by material (dust and/or electrons) distributed around the AGN
will make a larger contribution to the total observed flux than
if, as inferred for optically selected QSOs, we have a direct,
unobstructed view to the
bright unpolarized nucleus.
Given an asymmetric distribution and/or illumination of the scattering
material, high polarization can result.

The 2MASS AGN survey presents a large, well-defined
sample of QSOs that invites comparisons with
AGN selected by other means.
In this paper we present observations of 2MASS QSOs aimed at
determining their broad-band optical polarization properties.
We test the conclusion reached by \citet{cutri01}, based on
the near-IR colors of the sample, that some of the objects
in the general low-$z\/$ QSO population found by 2MASS
are obscured along our line of sight.
In addition, the polarization distribution of the new near-IR--selected QSOs
is compared to three largely radio-quiet QSO samples
(PG, BALQSO, and {\sl IRAS\/}
HIGs) for which polarization data are available.

\section{Observations}

The objects selected for observation were 89 spectroscopically-confirmed 
AGN from \citet{cutri01}. 
Of these, 70 objects meet the criteria that we adopt for 2MASS ``red'' QSOs. 
That is, $J-K_s > 2$ and $M_{K_s} < -23$.
Absolute magnitudes are determined assuming
$H_0 = 75$~km~s$^{-1}$~Mpc$^{-1}$,
$q_0 = 0$, and $\Lambda = 0$ which are used throughout this paper,
and are K-corrected using the near-IR spectral indices 
derived from the {\sl JHK$_s\/$} 2MASS magnitudes.
The $K_s\/$ luminosity criterion adopted for inclusion into the sample
was chosen to encompass the range of $M_{K_s}\/$ shown by the PG QSOs.
Only three 2MASS QSOs in this sample have $M_{K_s} > -25$, reflecting
the fact that the PG QSO sample includes many more objects with
$z < 0.1$ than the 2MASS sample.
In addition, all 2MASS AGN identified by \citet{cutri01} have
$\mid b \mid$ $>$ 30\dgr .
The sample of 2MASS QSOs and their optical polarization measurements are listed
in Table~1.

Table~2 lists 18 additional AGN and one unidentified object
observed in the polarization survey.
Most of these objects do not meet the $J-K_s > 2$ criterion, though their
absolute $K_s\/$ magnitudes place them in the range adopted for QSOs.
2MASSI J004125.3+134335 and 2MASSI J145744.9+202809 (2M004125 and 2M145744;
hereafter objects will be designated as 2M{\it hhmmss\/} in the text)
do not meet the 
$M_{K_s}\/$ criterion.
Mostly because of the relatively bright limiting apparent magnitudes
and color selection of the
2MASS AGN survey, all of the objects observed in the polarimetry program
have confirmed redshifts $<0.6$ except 2M092145 and
2M234905.
We have excluded the objects listed in Table~2 from the
analysis of the 2MASS QSO
sample so that we can solely concentrate on the polarization properties
of a well-defined low-redshift sample and allow for a more
straightforward comparison of these properties with other QSOs
in this redshift range.

\subsection{Photo-Electric Polarimetry}

A majority of the observations made during
1999 were obtained using the Two-Holer
Polarimeter/Photometer \citep{sitko85} on the Steward Observatory (SO)
1.5~m telescope atop Mt. Lemmon, AZ, or the Bok 2.3~m reflector on
Kitt Peak, AZ.
This instrument uses a semi-achromatic half-waveplate spinning at 20.65~Hz
to modulate incident polarization, and
a Wollaston prism to direct orthogonally polarized beams to two 
photomultiplier tubes.
All observations for this survey
were made unfiltered so that these generally optically faint objects could
be observed using the 1.5~m and 2.3~m telescopes.
Given the typically red optical colors of the QSOs and the location
of the H$\alpha$ line, the effective wavelength of unfiltered observations
usually falls within the
Kron-Cousins $R\/$ bandpass ($\lambda\lambda$6000--7000).
All observations were made with the smallest available circular aperture
(0.51~mm).
This subtends 4\farcs 3 diameter at the 
f/16 Cassegrain focus of the 1.5~m telescope.
For the f/9 Bok telescope, a lens is inserted before the aperture to slow
the beam to f/16 and the aperture translates to 2\farcs 9.
The lens has poor transmission at $\lambda < 4000$~\AA\ and therefore
unfiltered observations made at the Bok telescope will have slightly
redder effective wavelengths.
Sky subtraction is accomplished by moving the object out of the
aperture several times during an observation.

Data taking and reductions are thoroughly described in
\citet{moore87} and \citet{smith92}.
Uncertainties in the percent linear polarization ($\sigma_P\/$) and 
polarization position angle ($\sigma_\theta\/$,
where $\theta\/$ is the polarization position angle
measured in the usual astronomical
sense --- 0\dgr\ is North and 90\dgr\ is East)
quoted in Tables~1 and 2 are based on photon counting 
statistics.
The linear polarization measurements listed are not corrected for
statistical bias.
The polarization position angle is essentially undetermined when
$P/\sigma_P < 1.5$ \citep[see][]{wardle74},
and in these cases no value for $\theta\/$
is given.
Polarization position angles are calibrated using the grid of
polarization standard stars measured by \citet{schmidt92a}.
Unpolarized standard stars measured by \citet{schmidt92a} were occasionally
observed to check for instrumental polarization, which was found to be
$\lesssim 0.1$\%.
Observations made with Two-Holer are identified in the ``Comments''
column of Tables~1 and 2.

\subsection{CCD Imaging Polarimetry}

Observations of 2MASS AGN during 2000, along with a few 1999 observations, were
obtained using the CCD Imaging/Spectropolarimeter \citep{schmidt92b}
in its imaging mode at the Bok telescope.
The field of view for imaging polarimetry at the Bok telescope is
$51\arcsec \times 51\arcsec$ with an image scale of
$\sim$1.9~pixel/arcsecond.
Like Two-Holer, a semi-achromatic $\lambda$/2-waveplate is placed in front
of a Wollaston prism to analyze incident linear polarization.
Using a plane mirror in the grating location, the two
orthogonally polarized beams are imaged onto a thinned,
anti-reflection-coated 1200$\times$800 Loral CCD.
Four separate reads 
of the CCD chip are made with
the waveplate rotated through 16 positions to determine the linear Stokes
parameters, $Q\/$ and $U\/$, using
exposure times for each waveplate position of 30--200~s.
All observations were filtered using a KPNO
``nearly Mould'' $R\/$ filter ($\lambda\lambda$6000--7000).

Aperture photometry is performed on the co-added total 
flux images as
well as the $Q\/$ and $U\/$ flux images to extract the normalized
linear Stokes parameters for all usable targets in the field of view.
Tables~1 and 2 list the results using a circular digital aperture with a
diameter of 6 pixels ($\sim$3\farcs 2).
For consistency with surveys of other QSO samples, no attempt has been 
made to correct the polarization for starlight from the QSO host
galaxies that are apparent in many of the images (see \S3 and \S4).
There is no evidence for extended polarized emission, but this
does not strongly constrain the size of possible scattering regions
since the seeing/image quality of the imaging polarimetry over the 
long exposure sequences was typically
1\farcs 5--2\farcs 5 FWHM.
Values for $P\/$, $\theta\/$, and their uncertainties are presented in
the same manner as the Two-Holer data.

For 16 objects, observations were made on two or three nights
to improve on an initially low
signal-to-noise ratio (S/N)
measurement. 
Weighted means of the multiple observations are given in Table~2, though
in most cases, the average values are dominated by the higher S/N 
measurements resulting from the imaging polarimetry.
The multiple observations do not reveal any
polarization variations larger than twice the
observational uncertainty, and
we use the weighted means in the subsequent
analysis.

\subsection{Galactic Interstellar Polarization}

For a high galactic latitude sample of objects, such as the 2MASS-selected
AGN,
polarization due to transmission through aligned dust grains within 
the Milky Way is minimal.
\citet{impey90} and \citet{berriman90} both show that observing
a level of interstellar polarization ($P_{\rm ISP}\/$) higher than 0.6\%
is rare among objects with $\mid b \mid$ $>$ 30\dgr .
We observed at least one star or galaxy in the fields of 44 of the target 
AGN.
For those cases where $P_{\rm ISP}/\sigma_{P_{\rm ISP}}\/ \ge 2$,
we find evidence for $P_{\rm ISP} \ge 0.6$\% in only 10 fields.
These are summarized in Table~3.
Most of these 10 measurements come from the region 120\dgr$ < l
< 180$\dgr\ and close to the galactic latitude limit for the survey 
($-$30\dgr$ > b > -35$\dgr ).
No measurement of Galactic interstellar polarization with
$P_{\rm ISP}/\sigma_{P_{\rm ISP}}\/ \ge 3$
yields $P_{\rm ISP} \gtrsim 1.7$\%.

Comparison of the interstellar polarization measurements with the polarimetry
of the AGN suggests that the polarization of only 2M032458 and
2M095504 could be solely due to dust within our own
Galaxy.
At the same time, the intrinsic polarization of only 2M005010
appears to be significantly suppressed
by a local interstellar component that is orthogonal to the
measured polarization of the object.
Correcting for the interstellar polarization measured in this field,
the $R\/$-band polarization of 2M005010 is
$P = 4.7 \pm 1.2$\% (not corrected for statistical bias) at
$\theta = 50$\dgr$ \pm 7$\dgr .
Because the average uncertainty in $P\/$ for the entire
sample, $\sim$0.5\%, is similar to the 
highest levels of $P_{\rm ISP}\/$ that are
encountered, the observed polarization
of the 2MASS objects have not been adjusted for the
presence of Galactic interstellar
polarization.

\section{The Polarization of 2MASS QSOs}

\subsection{The Distribution of Polarization}

The distribution of linear polarization for the 70 QSOs is shown in
Figure~1.
The unfiltered and $R\/$-band measurements have been combined to construct
the histogram, and lacking multi-color polarization information,
no adjustments to the observed polarizations have
been made for differences between the effective bandpasses of the
Two-Holer and imaging polarimetry.

The optical broad-band polarimetry shows that the 2MASS QSOs are
a generally highly polarized sample of AGN.
More than 10\% of the sample (nine of 70 objects) have polarization above
3\% and a maximum polarization of $\sim$11\% is observed.
The results
can be contrasted to the unfiltered optical polarization of the
PG QSO sample \citep{berriman90} that is
shown in the lower panel of Figure~1.
For the PG sample, we only include the objects with $z < 0.6$ to match
the redshift range for the two QSO samples.
The average redshift of the 2MASS sample is $\langle z \rangle = 0.25$
(median = 0.24), whereas $\langle z \rangle = 0.19$ (median = 0.16) for
the 75 PG QSOs with $z < 0.6$.
Inclusion of the 21 higher-redshift PG QSOs does not change the obvious 
difference between the general polarization levels of the two samples,
since the highest polarization observed in the entire PG catalog is 
$P \sim 2.5$\% for
PG~1114+445 ($z = 0.144$) and PG~1425+267 ($z = 0.366$). 
For the purposes of comparisons
between the PG and 2MASS QSOs, the PG sample will
be limited to the 75 QSOs with $z < 0.6$ and $M_K < -23$.

The distributions of optical polarization show a dramatic difference between
optically selected QSOs represented by the PG sample, and
near-IR selected AGN having similar $K_s\/$-band luminosities
($-26 > M_{K_s} > -29$).
The high polarization of the 2MASS QSOs relative to UV/optically
selected QSOs supports the conclusion that
the red near-IR colors and high near-IR--to--optical flux ratios
arise from some degree of obscuration by dust in our line of sight.
Extinction and reddening by dust bias these objects against discovery
in traditional optical (UV-excess) AGN surveys.
High polarization is expected from a reddened, partially
obscured QSO population either because of the transmission of AGN light
through aligned dust grains (the interstellar polarization
mechanism) within the AGN host galaxy, or the
scattering of nuclear light by material
distributed fairly close to the AGN.
Optical spectropolarimetric results for several 
highly polarized 2MASS QSOs support the view that scattering of nuclear light
in regions near the nucleus is indeed the dominant polarizing mechanism
\citep{smith00a,smith00b}, and not dichroic absorption by aligned
dust grains.
Polarized light from
analogous scattering regions in UV/optically selected QSOs appears to
be completely swamped by the unpolarized, direct light from the nucleus,
leading to the low polarization observed for most QSOs measured in
past surveys \citep[e.g.,][]{stockman84,berriman90}.
We further explore this conclusion in the following sections.

\subsection{Polarization and Color}

If dust obscuration in our line of sight
plays a role in producing the high optical polarization of
2MASS QSOs relative to presumably unobscured QSOs found optically,
then a correlation might be expected between polarization and
color in the sense that redder, more obscured objects would tend
to be highly polarized.
Figures~2 and 3 show the observed polarization as a function of
$J - K_s\/$ and $B - K_s\/$ color indices, respectively.
The PG QSOs with $z < 0.6$ are also plotted in both figures 
using data from \citet{neugebauer87} and \citet{berriman90}.
It is readily apparent, especially with respect to $B - K_s\/$, that the
likelihood of detecting $P > 3$\% is higher for redder objects.
A non-parametric correlation test (Kendall's tau) applied to 2MASS sample
confirms the trends suggested in Figures~2 and 3, and these results are
summarized in Table~4.
The probability of finding a correlation this strong
between polarization
and $J - K_s\/$ or $B - K_s\/$ if these parameters are uncorrelated
is $< 0.5$\%.

High polarization is observed over the range of $2 < J - K_s\/ < 3$
spanned by the 2MASS sample, though most 2MASS QSOs with
$J - K_s \lesssim 2.4$ have $P < 3$\%.
The three objects with $J - K_s\/ > 3$ are also not highly polarized.
It is interesting to note that the near-IR color criterion for selection into
the
2MASS sample seems to be near something of a 
threshold for observing high polarization.
Three of the nine highly polarized 2MASS QSOs are at the red end of the
$J - K_s\/$ range spanned by the PG sample ($1 < J - K_s < 2.2$),
although none of the 12 PG QSOs that meet the 2MASS color selection
criterion have $P > 1$\%.

The $B - K_s\/$ color index is much more sensitive to the presence of
dust in our line of sight to the AGN
than $J - K_s\/$, and as expected, the 2MASS QSOs span a 
range of at least 5.7 magnitudes in $B - K_s\/$.
Four objects are
undetected in the blue Palomar Digital Sky Survey (DSS).
This is in stark contrast to the low-$z\/$ PG QSOs that span only about
2 magnitudes in $B - K_s\/$.
All QSOs with $P > 3$\% have $B - K_s > 5$ 
with the exception of 2M151653 ($B - K_s \sim 4.4$; $P \sim 9$\%).
If the 2MASS QSOs are similar to PG QSOs except that they are obscured 
by dust, high polarization objects appear in the sample
when there is $\gtrsim$1.5~mag of visual extinction of the direct 
nuclear light above that experienced by an
average PG QSO.

\subsection{Polarization and Near-IR Luminosity}

A strong trend is observed among the 2MASS QSOs between optical
polarization and the near-IR luminosity inferred from the 
apparent $K_s\/$ magnitudes (Figure~4 and Table~4).
Six of the 10 most luminous objects at 2.2$\mu$m are highly polarized, and
all objects with $P > 3$\% have $M_{K_s} < -26.5$. 
There are two probable effects that might preferentially
select intrinsically luminous QSOs in this sample
to be highly
polarized.
First, under the assumption that 2.2$\mu$m brightness is a more
accurate measure of intrinsic luminosity than optical brightness,
the most luminous objects are better able to illuminate
scattering regions distributed around the nuclear source.
Brighter and/or more extensive scattering regions around the nucleus
will generally result in more polarized, scattered light that
contributes a larger 
fraction to the total observed flux.
The second effect is the suppression of polarization
by unpolarized sources of optical light such as
starlight from
the host galaxy.
The relative contribution of the host galaxy to the total flux
is typically higher for lower luminosity AGN, and therefore,
the observed polarization is lower.

The 2MASS AGN sample has no morphological selection criteria and
inspection of the DSS and frames obtained from the imaging polarimetry 
easily reveal host galaxies for many of the 2MASS QSOs.
In several cases, the extended galactic emission outshines the QSO in the
optical.
For instance, only $\sim$50\% of the $R\/$-band flux measured
in a 20\arcsec\/-diameter aperture for 2M022150 is included in a 6\arcsec\/
aperture. 
By comparison, sophisticated optical observational techniques or near-IR
imaging have typically been required
to detect the host galaxies of UV/optically selected QSOs
\citep[see e.g.,][]{mcleod94,bahcall97}.
Clearly, the effect on the measured optical polarization
by host galaxy starlight
will tend to be greater for the 2MASS sample
than it is for other QSO samples.
This issue 
is discussed further in \S3.4 and \S4.

\subsection{Polarization and Optical Spectral Type}

One of the hallmarks of the 2MASS color-selected sample is the wide variety of
AGN types revealed by optical spectroscopy \citep{cutri01}.
The full range of AGN emission-line spectral types
are represented in the 70 2MASS QSOs, from Type 1 to Type 2/LINER.
A majority of the objects show some evidence for broad emission lines
(Type~1--1.9) with only 9 objects not showing a broad component in their
Balmer line profiles.

We have divided the QSO sample into categories using the 
spectral classifications determined by \citet{cutri01}:
Type 1 (22 objects; which includes Type~1.2 objects),
Type~1.5 (19 objects), and Type~1.8--2/LINER/Starburst (15 objects).
Objects that show broad H$\alpha\/$ emission lines, but whose
available spectra either do not include H$\beta\/$ (``Type~1.b''; 3 objects),
or have no detected H$\beta\/$ emission line 
(``Type~1.x''; 10 objects) have been combined into a
fourth category (Type~1.?).
Spectral classification has been ambiguous for 2M232745 and we have
included it with the Type~1.x and 1.b objects.
The properties of these categories and the $z < 0.6$ PG QSOs
are summarized in Table~5.

Inspection of Figures~2, 3, or 4 reveals that all categories,
with the exception of Type~1.?,
are represented among the nine highly polarized examples.
Over one quarter of the 19 QSOs classified as Type~1.5 are
highly polarized.
In the case of 2M165939, the polarized flux
spectrum shows a blue continuum, broad Balmer emission lines, and that
the NLR features are unpolarized \citep{smith00a}.
Although there is a broad emission-line component in the 
total flux spectrum, these polarization properties imply
that the AGN is partially
hidden from our direct view.
Moreover,
the scattering material must lie further away from the continuum source
than the broad emission-line
region (BLR), but be
close to, or
within the NLR.

The relative lack of high polarization among Type~1 objects (two
of 22 objects) is reminiscent of the $z < 0.6$ PG sample that is
composed of Type~1--1.5 objects (adopting the classifications 
compiled by \citealt{veron01}), where none have $P > 3$\%.
Although 2MASS detects redder Type~1 objects, many that are presumably
reddened by dust, direct unpolarized light from the nucleus
is apparently still strong enough to swamp
any scattered, polarized light sources.
Alternatively, as with any of the low-polarization objects, the scattered
light component could also be reddened, keeping its contribution to 
total observed flux too small to yield $P > 3$\%.
One of the two highly polarized Type~1 QSOs,
2M151653, has been shown by 
\citet{smith00a} 
to be polarized in the same manner as 2M165939.
That is, material close to or within the NLR scatters light
into our line of sight
from an active nucleus partially obscured from our
direct view.

All of the highly polarized QSOs discovered in this survey exhibit
evidence for a broad-line region in their total flux spectra, albeit
weakly in the cases of 2M010835 (Type~1.9; $M_{K_s} = -27.6$) and
2M171559 (Type~1.8; $M_{K_s} = -28.1$).
Therefore,
the high polarization of the 2MASS sample is not
the result of the inclusion of Type~2 objects with 
hidden continuum sources and BLRs
like those that exist in several highly polarized
Seyfert~2 nuclei \citep[see e.g.,][]{tran95b}.

On the other hand, the two highly polarized narrow emission line-dominated
objects are the most luminous objects of their type in the sample.
No other Type~1.8--2 objects have $M_{K_s}\/ < -27$ and
all but 2M145410, 2M163736, and 2M230304 have $M_{K_s}\/ > -26$.
Within the context of the unified AGN scheme, Type~2 objects are
thought to be more
heavily obscured by dust than Type~1s.
This naturally explains why the $K_s\/$ luminosity 
of the Type~2 objects is generally lower 
than that of Type~1--1.5 and why $B - K_s\/$ is typically larger for Type~2s
(Table~5).
The prevalence of intermediate-type QSOs among the highly polarized 2MASS
QSOs suggests that unpolarized light sources dilute the optical
polarization of objects of more extreme spectral type.
Type~2 objects may exhibit low polarization
because of the relatively large flux contribution from stars in the
host galaxy,
assuming that there is no large systematic difference in
host galaxy luminosity between types,
or from other unpolarized sources such as the featureless
continuum (FC2) that is argued to be present in the spectra of some Seyfert~2
nuclei \citep[e.g.,][]{tran95c}.

The effect of the host galaxy on the polarized AGN flux can be
roughly estimated for the different spectral types of 2MASS QSOs
summarized in Table~5.
Given the median luminosity and colors of the Type~1--1.5 objects, the observed 
$R\/$-band polarization is $\sim$0.8$P_0\/$, where $P_0\/$ is the intrinsic
$R\/$-band polarization of the AGN, assuming that
the host galaxy is similar to the
average host found for UV-excess QSOs
\citep[$M_R \sim -21.9$ for radio-quiet QSOs;][]{bahcall97}.
In comparison, a similar host for a typical Type~1.8--2 2MASS QSO will dilute
$P_0\/$ by about a factor of 2.

If the 2MASS Type~2 QSOs are high-luminosity analogs to Seyfert~2 nuclei,
it is not surprising that high broad-band optical ($R\/$) polarization is
not observed in these objects.
Even the archetype of Seyfert~2 nuclei that harbor Type~1 polarized spectra,
NGC~1068, does not show $P > 3$\% in the red
\citep{angel76,miller83}.
Much of the strong increase in polarization in the blue in Seyfert~2s is
attributed to the decreased dilution by starlight
\citep[see e.g.,][]{antonucci85,tran95a}.
The 2MASS QSO polarization measurements have an effective wavelength
$>$6000~\AA , and the redshifts of the objects in the sample are not
high enough that the polarimetry includes much flux blueward of the
Ca~H and K break where the starlight contribution is diminished.
In addition, any reasonable aperture
will include a majority of the flux
from the host galaxy for a typical 2MASS QSO, whereas
much of the host galaxy starlight is avoided
using similar apertures for Seyfert~2 galaxies simply because of
their much lower redshifts.

Finally, there are no highly polarized examples among the objects placed
in the Type~1.? category.
This category is dominated (10 of 14 objects) by QSOs that show no evidence
for H$\beta\/$ emission in their confirmation spectra, though broad
H$\alpha\/$ is observed \citep{cutri01}.
The objects have a median $M_{K_s}\/$ similar to Types~1
and 1.5 QSOs in the 2MASS sample, but their $B - K_s\/$
color indices are more typically like the much redder Type~1.8--2 objects.
The red colors and large Balmer decrements suggest that we are not 
seeing much AGN light at optical wavelengths and, as with the Type~1.8--2
objects, the host galaxy dilutes the polarized flux.
The low optical polarization of the highly reddened Type~1.x QSOs
could also be the result of dust obscuring the scattering regions.

\clearpage

\section{Discussion}

\subsection{Radio and Infrared Properties}

The optical broad-band polarization results
are consistent with the interpretation that the 2MASS sample includes
a fraction 
of low-$z\/$, radio-quiet QSOs that are obscured along
our line of sight
\citep{cutri01}.
In \S3 the 2MASS
QSOs were compared with the predominantly radio-quiet
PG QSOs in the same
redshift range.
This comparison is motivated not only by the similar numbers of
objects and the high S/N polarimetry
that is available,
but also because the two 
samples have similar luminosity distributions at $K_s\/$.
However, it can be seen in Figure~4 that the 2MASS sample lacks objects 
as bright at $K_s\/$ as the most luminous PG QSOs.
\citet{cutri01} suggest that this is caused by selection
biases in the 2MASS AGN sample such as the exclusion of known AGN, and that the
extinction necessary to achieve $J - K_s \ge 2.0$
essentially rules out finding new objects as bright as the most luminous
unobscured QSOs.
This should not be the case in the mid- to far-IR where
emission from dust near the nucleus will dominate the observed flux.
An important test of the
equivalence of the 2MASS and optically-selected QSOs is a comparison
of their spectral energy distributions (SEDs)
from mid-IR to radio wavelengths where orientation
effects are mitigated.

Nearly half (33 of 70) of the 2MASS QSOs are detected at 1.4~GHz by either the
NRAO VLA Sky Survey
\citep[NVSS;][]{condon98} or
the VLA Faint Images of the Radio Sky at Twenty-Centimeters (FIRST) survey
\citep{becker95}, and about half (5 out
of 9) of the
highly polarized QSOs are detected.
The detection rate is about the same for the PG QSOs with $z < 0.6$
(38 of 75 objects),
even though the 2MASS sample by design
omits known AGN in radio catalogs.
No object has a 1.4~GHz radio power 
(P$_{1.4 \rm{GHz}}\/$)
above $10^{25}$~W~Hz$^{-1}$, while the PG sample boasts 10 objects
with $25 < \log {\rm P_{1.4 \rm{GHz}}} < 27.5$ (Figure~5).
In addition to having a number of powerful radio sources, the radio-detected
PG sample
includes 14 objects with low radio luminosity
($\log {\rm P_{1.4 GHz}} < 23$).
Eleven of these objects have $z < 0.1$.
The low-luminosity objects are missing from the 2MASS sample primarily
because of its higher average redshift.
Thus, although
selection biases strongly affect the 2MASS sample,
the 2MASS QSOs are roughly consistent in radio
luminosity with the radio-quiet PG QSOs.
There is no evidence for any relationship between radio luminosity
and high optical polarization among the 2MASS objects.

Despite the similarity in the overall
1.4~GHz detection statistics for the 2MASS and PG
QSO samples, the Type~1 2MASS objects are under-represented in the 
radio surveys.
Only three of 22 Type~1 QSOs are detected by FIRST or NVSS,
whereas $\sim$60\% of objects in the other 2MASS spectral categories are
detected at 1.4~GHz (Table~5).
The reason behind the lack of radio sources associated with Type~1 objects 
in the 
near-IR selected sample compared to the optically selected PG QSOs is
not readily apparent, but this discrepancy hints at possible
intrinsic differences in the central engine and/or AGN environment between the
two samples.

Comparison between the 2MASS and PG QSO
samples in the mid infrared is difficult because of the
small number of objects detected by
{\sl IRAS\/}.
Only 15 of the 70 2MASS QSOs are detected at 60$\mu$m in the {\sl IRAS\/}
Faint Source Catalog (FSC).
Nineteen PG QSOs with $z < 0.6$ were detected and the slightly higher success
rate is due to the  
PG sample containing more objects with $z < 0.1$ that are
more likely to be bright enough to be seen by {\sl IRAS\/}.
Figure~6 shows that the {\sl IRAS\/}-detected 2MASS QSOs are generally
more luminous
at 60$\mu$m for a given near-IR luminosity than 
the PG QSOs, but this is at least
partially a selection bias caused by the higher redshift of the 2MASS sample.
Two highly polarized 2MASS QSOs were detected by {\sl IRAS\/} and neither
of these objects show an excess of 60$\mu$m emission above that expected
from a PG QSO.

As with the radio surveys, the Type~1 2MASS QSOs are rare
in the sample detected at 60$\mu$m, with
only 2M171442 being represented.
The remainder of the 60$\mu$m detections summarized in Table~5
are distributed among Types~1.5
(6 of 19 objects), 1.x (6 of 10 objects), 1.8 (2M022150), and
1.b (2M130700).
The large representation of Type~1.x QSOs in Figure~6 is striking.
These objects tend to be the most luminous QSOs at 60$\mu$m 
for a given $M_{K_s}\/$.
It may be that these systems contain more dust than other types of
QSOs, a difference that could account for
the observed infrared emission, redder optical and near-IR
colors, and large Balmer decrements.
One possibility is that the excess dust has been added via a merger with
another galaxy.
This would also help to explain the low polarizations observed for the
Type~1.x objects since a large amount of dust in the system could
obscure the scattering regions.
The presence of two stellar systems would also help to dilute any polarized
flux from the AGN.
Future high-resolution optical and near-IR imaging of the 2MASS sample
can explore the possible role that galaxy mergers play in QSOs.

The high sensitivity of the {\it Space Infrared Telescope Facility\/}
({\sl SIRTF\/}) is well suited to examine the mid-IR SEDs of
AGN samples.
Future {\sl SIRTF\/} observations 
may resolve the issue of whether near-IR selected, red QSOs
are similar to optically-selected QSOs in their mid- and far-IR
properties as expected if orientation
to our line of sight is indeed the major discriminant
between AGN types at shorter wavelengths, or are systematically more luminous
as suggested by the few {\sl IRAS\/} detections.

\subsection{Comparison with Other QSO Samples}

Broad-band optical polarization surveys exist for the
BALQSOs and {\sl IRAS\/} HIGs
that help place the new 2MASS sample in context with current ideas concerning
AGN.
In Figure~7 we plot the cumulative distributions of optical polarization
for these two samples along with those of the 2MASS and PG QSOs.
Results of Kolomagorov-Shmirnov (KS) tests on the distributions are given
in Table~6, where it can be seen that the observed differences
between the PG QSOs and the other three samples are far larger
than would be expected
if the samples were
drawn from the same parent population.

\citet{schmidt99} have shown that the polarization distributions for the 
PG QSOs and BALQSOs can be made similar
if $P\/$ for the PG sample
is multiplied by $\sim$2.4.
That is, the higher polarization observed in BALQSOs can be explained
if the direct, unpolarized nuclear light from these objects
is attenuated in our line of sight
by $\sim$1~mag above that experienced by a typical PG QSO.
This interpretation is in keeping with evidence that low-ionization BALQSOs are
reddened by dust \citep{sprayberry92} and with detailed spectropolarimetry
that implies a covering factor $< 1$ for the absorbing
gas and material partially
obscuring the nucleus \citep[e.g.,][]{glenn94,
hines95a,goodmill95,cohen95,goodrich97}.
The implications
of the broad-band polarimetry
are not changed by the addition of the observations 
of \citet{hutsemekers98} and \citet{lamy00}
that increases the BALQSO sample to
85 objects.

The cumulative distributions of the 2MASS QSOs and BALQSOs
(Figure~7) follow each other closely for low polarization levels ($P < 1$\%). 
\citet{schmidt99} have shown that the
larger statistical bias associated with the higher average observational
uncertainty of the fainter BALQSOs
does not explain the generally higher levels of polarization 
observed for BALQSOs relative to the PG sample for
$P \sim 1$--2\%.
Because the mean observational uncertainties
of the 2MASS QSO and BALQSO polarization measurements are similar,
the divergence of the 2MASS
and PG QSO distributions at low polarization levels also cannot be an artifact
of statistical bias in the measurements.
Therefore, the 2MASS QSO sample
is not distinguished
from the PG QSO sample simply by the addition of a 
small fraction ($\sim$10--15\%) of highly polarized objects.
Like the BALQSOs, the 2MASS QSO sample {\it as a whole\/} exhibits
higher polarization than the PG QSOs.

The congruence of the 2MASS QSO and BALQSO distributions
at low polarization levels suggests
a possible relationship between these two samples.
In fact, despite a divergence at high polarization,
a KS test yields a reasonable probability that the 
samples could be drawn from the same parent population
(P$_{\rm KS} \sim 0.2$).
Of the four known high-redshift ($z > 1$) AGN found by 2MASS, one 
shows broad Mg~II absorption \citep{cutri01}.
It is possible that the 2MASS QSO sample is made up in part of objects that are
low-redshift analogs to the high-$z\/$ BALQSOs, but there are
some effects that must be taken into account to properly compare
the polarization properties of the two samples.

First, the BALQSOs tend to be at a much higher redshift
($\langle z \rangle = 1.8$) than the 2MASS QSOs ($\langle z \rangle = 0.25$).
As a result,
optical polarimetry of most BALQSOs measures $P\/$ in the rest-frame
ultraviolet.
Any general wavelength dependence in the continuum polarization between
the optical and near-UV will result in a systematic difference between the 
samples.
Possibly more important is the effect of host galaxy starlight.
Unlike the 2MASS objects,
polarization measurements of BALQSOs are unlikely to be affected by
the host galaxy
because of the overwhelming brightness of the nucleus
and the fact that
the observations are made blueward of the Ca~H and K break where the flux
from galaxies is diminished.

Correcting the polarimetry of 2MASS QSOs for host galaxy
starlight will result in a greater differentiation between the
polarization distributions of the 2MASS and BALQSO samples.
We have attempted to measure the effect on $P\/$ of the
host galaxy in several 2MASS QSOs using the imaging polarimetry data.
Various digital apertures were tried to explore the dependence of $P\/$
on aperture size.
However, with an image scale of $\sim$0\farcs 5 pixel$^{-1}$,
even the smallest practical apertures admit a large
fraction of the total stellar flux,
and in no case is the expected
trend of decreasing $P\/$ with increasing aperture size unambiguously
observed.

A final difference to be noted between the polarization properties of
2MASS QSOs and BALQSOs is presented by spectropolarimetry.
\citet{schmidt99} summarize the general spectropolarimetric properties
found for BALQSOs from various studies which imply that scattering
produces the observed polarization \citep[see also,][]{ogle99}.
In most cases, the broad emission lines are not as highly
polarized as the continuum, and in several instances there is no 
evidence for broad emission lines in the polarized flux spectrum at all.
Spectropolarimetry of highly polarized 2MASS QSOs, on the other hand,
shows that the broad
Balmer lines tend to be polarized to about the same level as the continuum
\citep{smith00a,smith00b}.
This is consistent with the view that the continuum source is
more heavily obscured than the BLR in BALQSOs, but the two
are attenuated by about the same amount in the 2MASS objects.

Future ultraviolet spectroscopy will provide a direct test of the possible
relationships between the 2MASS QSOs and BALQSOs.
Determining the fraction of near-IR selected QSOs that show broad absorption
lines and comparing the UV properties of
2MASS QSOs with other samples will likely yield important information
about the structure of luminous, obscured AGN and about
the role that orientation plays in the selection and identification of these
objects.
In a similar vein, soft X-ray observations of 2MASS QSOs can
measure the X-ray absorption in the
line of sight and these values can be compared
to the absorbing column densities
exhibited by other AGN samples 
\citep[see e.g.,][]{mathur95,brandt00,gallagher01,green01}.
Results from the {\it Chandra\/} X-ray Observatory suggest
that 2MASS QSOs exhibit column densities,
$N_{\rm H} \sim 10^{21-23}$~cm$^{-2}$
\citep{wilkes01}.

The HIGs are another important QSO sample \citep{cutri94}
that may be closely related to
objects found from their near-IR colors.
Like the 2MASS sample, the HIGs are composed of AGN of various
optical spectral types and show larger infrared-to-optical
flux ratios than are observed for UV/optically selected
QSOs.
\citet{hines94} and \citet{wills97} have found that the {\sl IRAS\/}
HIGs as a class are optically highly polarized (see Figure~7).
Spectropolarimetry of the most luminous HIGs with Type 2-like spectra
has revealed that these objects are powered by QSOs obscured in our line 
of sight
\citep{hines95b,goodrich96}.

The small {\sl IRAS\/} sample occupies a larger region of redshift
space ($z = 0.04$--2.3) than do the 2MASS QSOs ($z < 0.6$), and it is the 
high-redshift (most luminous)
HIGs that tend to be the most highly polarized.
This resembles the trend between $P\/$ and near-IR
luminosity found for the 2MASS QSOs (\S3.3).
Another parallel between the two samples is that objects
with the highest infrared-to-optical flux ratios (as measured by
$F_{K_s}\//F_B\/$
for the 2MASS sample and $F_{60 \mu {\rm m}}\//F_V\/$ for the HIGs) tend to be
the most highly polarized.
These trends in the polarization properties suggest
that the HIGs may represent the high-luminosity tail of the
dust-obscured QSO population.
The Type 2 {\sl IRAS\/} HIGs are among
the most luminous objects known and are highly polarized
\citep[see e.g.,][]{hines95b,hines99}, whereas
the lower luminosity Type 2 2MASS QSOs do not show high optical
broad-band polarization.
We ascribe this difference to the dominance
of the scattered light over host galaxy starlight in the HIGs, and
to the fact that
the polarimetry of the 2MASS QSOs does not measure the polarization
in the rest-frame UV.

\section{Summary and Conclusions}

The broad-band optical polarization of QSOs selected by their near-IR
colors are consistent with the view
that many of these objects are obscured to some degree by dust.
Over 10\% of a sample of 70 ``red'' QSOs discovered by 2MASS \citep{cutri01}  
shows high polarization ($P > 3$\%), and values range up to $\sim$11\%.
These high levels of polarization are not present among objects in
standard UV-excess catalogs of
QSOs except for OVVs and a small number of BALQSOs.
The polarization of 2MASS QSOs is correlated with near-IR
luminosity and color, and with near-IR--to--optical flux ratio in the sense
that more luminous and redder objects tend to show high polarization.
These trends suggest that the polarization arises from the scattering
of nuclear light by material located close to a partially obscured AGN.
Spectropolarimetry of two highly polarized 2MASS QSOs confirms that 
scattering is the dominant polarizing mechanism
in these objects with the scattering material
located close to, or within, the NLR \citep{smith00a}.
Determining the nature of
the scattering material (either dust and/or electrons),
however, likely awaits careful modeling of high-quality spectropolarimetry
at UV, optical, and infrared wavelengths \citep[see e.g.,][]{hines01}.

The near-IR color selection of the 2MASS AGN survey is unbiased with regard
to optical spectral type,
and we find that high polarization is found among
nearly all types, especially those
classified as Type~1.5--1.9 by \citet{cutri01}.
Like optically selected QSOs,
few of the broad emission line-dominated (Type~1) QSOs have $P > 3$\%.
This is presumably because the obscuration of direct AGN light is
sufficiently low that the relative contribution 
of any scattered component is small.
High polarization is not observed in Type~2 QSOs in this sample,
although their colors and $K_s\/$ luminosities 
are consistent with these objects being among the 
most heavily obscured in the sample.
Dilution of scattered, polarized light by starlight from the host galaxy is
suggested as the reason for low observed optical polarization in this case.
This assertion is consistent with the fact that most Seyfert~2 galaxies that are
highly polarized in the UV show $P < 3$\% for $\lambda > 5000$~\AA . 
Ultraviolet polarimetry and high quality imaging provide
obvious tests of the effect of host galaxy starlight on the polarization
of 2MASS QSOs.

The distribution of optical polarization for the 2MASS QSOs
further 
substantiates the trend that higher polarization
is seen in QSO samples that have higher
apparent IR--to--optical luminosity ratios
($L_{\rm IR}\//L_{\rm opt}\/$).
The 2MASS objects are not as extreme in $L_{\rm IR}\//L_{\rm opt}\/$ as the
{\sl IRAS\/} HIGs, and neither is their general level of polarization.
On the other hand, the 2MASS
sample is much more highly polarized than the PG QSOs,
which generally have the lowest $L_{\rm IR}\//L_{\rm opt}\/$.
This is presumably
because our view of the nuclear region is not hindered by dust in
UV-excess AGN.
Although the 2MASS and BALQSO distributions in $P\/$ are fairly
similar, corrections
for starlight and for differing observing bandpasses
are likely to
better differentiate these two samples and thereby increase the significance
of the correlation between $P\/$ and $L_{\rm IR}\//L_{\rm opt}\/$.
This correlation implies that orientation effects and/or the dust
covering factor are largely responsible
for differences in the optical and IR properties between various samples
of high-luminosity AGN. 
Alternatively, the near-IR--selected QSOs
may represent a different 
evolutionary stage than the optically selected, UV-excess QSOs.
Further orientation-independent investigations of these
samples, such as imaging of host galaxies and environments, extended
radio emission, and IR observations using {\sl SIRTF\/} will be necessary
to distinguish between evolutionary and orientation effects.

Finally,
the polarization properties of the 2MASS QSOs suggests
that this new near-IR--selected sample may include a large proportion
of low-redshift BALQSOs.
Ultraviolet observations can directly test this possibility, and
together with X-ray measurements, investigate the nature of the
obscuring material.

\acknowledgments

We thank George Rieke and Frank Low for useful discussions and the 
National Aeronautics and Space Administration (NASA)
and the Jet Propulsion Laboratory (JPL) for 
support through {\sl SIRTF\/}/MIPS and Science Working Group contracts 960785
and 959969 to The University of Arizona.
RMC and BON acknowledge the support of the JPL
operated by the California Institute of Technology under contract to NASA.
Polarimetric instrumentation at Steward Observatory is maintained, in part,
through support by National Science Foundation (NSF) grant AST 97--30792.
This publication makes use of data products from the Two Micron All Sky Survey, which is a joint project of the University of Massachusetts
and the Infrared Processing and Analysis
Center/California Institute of Technology,
funded by NASA
and the NSF.
The DSS is based on photographic data of the National Geographic
Society--Palomar Observatory Sky Survey obtained using the Oschin Telescope
on Palomar Mountain and was produced at the Space Telescope Science Institute
under US Government grant NAG~W-2166.
Luke Moore assisted with preliminary analysis of the initial observations.

\clearpage
\oddsidemargin=-0.30in
\evensidemargin=-0.30in

\begin{deluxetable}{lrcrrrrcrrrrc}
\tablecolumns{13}
\tablewidth{0pc}
\tabletypesize{\small}
\tablecaption{Optical Polarimetry of 2MASS Red QSOs}
\tablehead{
\colhead{Object}  &
\colhead{$z\/$\tablenotemark{a}} & \colhead{Type\tablenotemark{a}} &
\colhead{$K_s\/$\tablenotemark{b}} & \colhead{$J-K_s\/$\tablenotemark{b}} &
\colhead{$B-K_s\/$\tablenotemark{b}} &
\colhead{$M_K\/$} &
\colhead{UT Date} &
\colhead{$P\/$\tablenotemark{c}} &
\colhead{$\sigma_P$} &
\colhead{$\theta\/$} &
\colhead{$\sigma_{\theta}\/$} &
\colhead{Comments\tablenotemark{d}} \\
\colhead{(2MASSI J)} & \colhead{} & \colhead{} & \colhead{} & \colhead{} &
\colhead{} &
\colhead{} &
\colhead{({\it yymmdd\/})} & \colhead{(\%)} &
\colhead{(\%)} & \colhead{(\dgr )} & \colhead{(\dgr )} & \colhead{}}
\startdata

000703.6+155423 & 0.114 & 1.8 & 13.10 & 2.13 & 4.30  & $-$25.42 & 991011  & 0.99 & 0.81  & \nodata & \nodata & 2H90 \\
000810.8+135452 & 0.185  & 2 & 14.40 & 2.10 & 5.70  & $-$25.29 & 000107  & 0.79 & 0.47 & 101.9 & 22.6 & \\
002614.5+184612 & 0.319 & 1 & 14.56 & 2.15 & 3.94  & $-$26.57 & 991012  & 1.01 & 1.27  & \nodata & \nodata & 2H90 \\
002924.5+242430 & 0.370 & 1 & 14.29 & 2.16 & 3.11  & $-$27.25 & 991012  & 1.70 & 0.88  & 89.2 & 18.5 & 2H90 \\
004118.7+281640 & 0.194 & 1 & 12.50 & 2.04 & 3.40  & $-$27.29 & 990909  & 2.16 & 0.27 & 103.5  & 3.5 & 2H60 \\
005010.1+280619 & 0.277 & S & 15.13 & 2.64 & $>$5.87  & $-$25.84 & 991016  & 2.28 & 0.75  & 49.2  & 9.5 & \\
005055.7+293328 & 0.136  & 2 & 13.22 & 2.11 & 5.68  & $-$25.72 & 990911  & 2.47 & 0.49  & 98.8  & 5.7 & 2H60 \\
 & & & & & & & & & & & & \\
010230.1+262337 & 0.194 & 1.x & 13.83 & 2.25 & 6.87  & $-$26.03 & 991016  & 2.11 & 0.61 & 120.5  & 8.3 & \\
010607.7+260334 & 0.411 & 1 & 14.61 & 2.69 & $>$6.39  & $-$27.58 & 991016  & 6.49 & 0.89 & 117.0  & 3.9 & \\
 & & & & & &   & 000108  & 5.79 & 0.69 & 112.2  & 3.4 & \\
 & & & & & & & ave.   &  6.03 & 0.54 & 114.1  & 2.6 & \\
010835.1+214818 & 0.285  & 1.9 & 13.46 & 2.75 & 6.54  & $-$27.64 & 990913  & 4.45 & 1.92 & 116.9 & 14.4 & 2H60 \\
 &   & & & & &   & 991011  & 5.93 & 1.15 & 119.7  & 5.6 & 2H90 \\
  &  & & & & & & ave.   &  5.45 & 0.99 & 119.1  & 5.2 & \\
012031.5+200327 & 0.087  & S & 12.47 & 3.85 & 5.63  & $-$25.66 & 990912  & 3.65 & 1.54 & 125.9 & 14.0 & 2H60 \\
  &  & & & & &   & 991012  & 0.74 & 1.67  & \nodata & \nodata & 2H90 \\
   & & & & & & & 000108  & 0.65 & 0.27 & 128.9 & 13.8 & \\
   & & & & & & & ave.   &  0.74 & 0.26 & 128.5 & 11.3 & \\
015721.0+171248 & 0.213 & 1.x & 13.16 & 2.70 & 7.34  & $-$27.09 & 991012  & 1.17 & 2.09  & \nodata & \nodata & 2H90 \\
   & & & & & &   & 000107  & 1.50 & 0.48  & 65.5  & 9.2 & \\
   & & & & & & & ave.   &  1.44 & 0.47  & 66.6  & 9.9 & \\
022150.6+132741 & 0.140 & 1.8 & 13.25 & 2.38 & 5.65  & $-$25.82 & 991011  & 1.06 & 1.72  & \nodata & \nodata & 2H90 \\
  &  & & & & &   & 000107  & 0.40 & 0.25  & 12.1 & 23.7 & \\
   & & & & & & & ave.   &  0.39 & 0.24  & 13.6 & 24.0 & \\
023430.6+243835 & 0.310 & 1.5 & 13.75 & 2.20 & $>$7.25  & $-$27.33 & 991011  & 2.57 & 0.46  & 99.1  & 5.1 & 2H90 \\
 & & & & & & & & & & & & \\
024807.3+145957 & 0.072 & 1 & 12.65 & 2.16 & 1.85  & $-$24.79 & 000108  & 0.37 & 0.15  & 75.7 & 13.6 & \\
032458.2+174849 & 0.328 & 1 & 12.82 & 2.37 & 4.18  & $-$28.49 & 990912  & 1.23 & 0.38 & 119.2  & 8.8 & 2H60 \\
034329.6+132519 & 0.314 & 1 & 14.28 & 2.12 & 3.42  & $-$26.79 & 991012  & 1.38 & 0.76  & 0.8 & 20.1 & 2H90 \\
034857.6+125547 & 0.210 & 1.x & 13.60 & 3.30 & 6.10  & $-$26.82 & 991016  & 2.23 & 0.73  & 69.5  & 9.4 & \\
081652.2+425829 & 0.235 & 1 & 13.73 & 2.11 & 3.57  & $-$26.57 & 991012  & 1.13 & 0.56 & 169.3 & 17.4 & 2H90 \\
082311.3+435318 & 0.182 & 1.5 & 12.85 & 2.08 & 3.65  & $-$26.80 & 991012  & 1.31 & 0.71  & 73.6 & 19.7 & 2H90 \\
085045.3+172003 & 0.343 & 1.5 & 14.46 & 2.22 & 3.74  & $-$26.91 & 000106  & 2.09 & 0.23 & 100.3  & 3.2 & \\
 & & & & & & & & & & & & \\
085116.8+120028 & 0.370 & 1 & 14.71 & 2.26 & 3.89  & $-$26.89 & 000106  & 0.61 & 0.36  & 93.3 & 22.1 & \\
091602.6+210617 & 0.422 & 1.5 & 14.99 & 2.06 & 3.32  & $-$26.86 & 000107  & 0.81 & 0.31  & 1.9 & 12.2 & \\
091848.6+211717 & 0.149 & 1.5 & 12.55 & 2.25 & 5.95  & $-$26.65 & 000107  & 6.30 & 0.14 & 154.1  & 0.6 & \\
092049.0+190320 & 0.156 & 1.b & 14.92 & 2.06 & 5.78  & $-$24.34 & 000107  & 1.15 & 0.52  & 19.9 & 15.5 & \\
092151.2+175855 & 0.356 & 1.5 & 14.79 & 2.06 & 2.71  & $-$26.59 & 000107  & 0.40 & 0.32  & \nodata & \nodata & \\ 
094636.4+205610 & 0.280 & 1.5 & 13.72 & 2.08 & 3.98  & $-$27.03 & 000107  & 0.22 & 0.24  & \nodata & \nodata & \\ 
094927.7+314110 & 0.308 & 1.x & 13.28 & 2.09 & 3.22  & $-$27.72 & 000107  & 0.44 & 0.16 & 148.8  & 12.0 & \\
 & & & & & & & & & & & & \\
095504.5+170556 & 0.139 & 1 & 13.44 & 2.03 & 3.96  & $-$25.54 & 000108  & 0.34 & 0.18  & 87.2 & 18.6 & \\
100121.1+215011 & 0.248 & S & 14.68 & 2.18 & 5.52  & $-$25.79 & 000505  & 2.12 & 0.98 & 138.0 & 15.7 & \\
  &   & & & & &   & 000509  & 0.95 & 0.62 & 124.7 & 25.1 & \\
  &  & & & & & & ave.   &  1.25 & 0.52 & 130.9 & 13.8 & \\
101400.4+194614 & 0.110 & 1.5 & 12.37 & 2.01 & 4.13  & $-$26.05 & 000108  & 0.67 & 0.12  & 98.7  & 5.2 & \\
101649.3+215435 & 0.257 & 1 & 13.94 & 2.03 & 3.16  & $-$26.56 & 000108  & 0.24 & 0.18  & \nodata & \nodata & \\ 
102724.9+121920 & 0.231 & 1.5 & 13.22 & 2.06 & 5.28  & $-$27.02 & 000107  & 1.83 & 0.31 & 141.2  & 4.9 & \\
105144.2+353930 & 0.158  & L & 13.54 & 2.11 & 5.06  & $-$25.77 & 000108  & 0.54 & 0.29  & 47.8 & 19.0 & \\
125807.4+232921 & 0.259 & 1 & 13.45 & 2.07 & 3.85  & $-$27.09 & 000108  & 1.22 & 0.13 & 105.8  & 3.0 & \\
 & & & & & & & & & & & & \\
130005.3+163214 & 0.080 & 1.x & 11.86 & 2.20 & 5.24  & $-$25.84 & 000108  & 1.68 & 0.14  & 43.9  & 2.3 & \\
130700.6+233805 & 0.275  & 1.b & 13.45 & 3.34 & 7.58  & $-$27.79 & 000505  & 1.58 & 0.96  & 29.1 & 23.1 & \\
  & & & & & & & 000508  & 3.18 & 0.83  & 40.9  & 7.5 & \\
  & & & & & & & ave.   &  2.45 & 0.63  & 37.7  & 7.3 & \\
132917.5+121340 & 0.203 & 1 & 14.12 & 2.02 & 4.58  & $-$25.78 & 000505  & 2.30 & 0.37  & 38.3  & 4.6 & \\
134915.2+220032 & 0.062 & 1.5 & 12.24 & 2.21 & 3.27  & $-$24.87 & 000505  & 2.88 & 0.22 & 105.4  & 2.2 & \\
140251.2+263117 & 0.187 & 1 & 12.67 & 2.11 & 3.83  & $-$27.05 & 000505  & 0.21 & 0.21  & \nodata & \nodata & \\ 
145331.5+135358 & 0.139 & 1.x & 13.09 & 2.23 & 4.51  & $-$25.93 & 000505  & 0.81 & 0.29  & 73.4 & 11.4 & \\
145406.6+195028 & 0.260 & 1.5 & 14.71 & 2.37 & 5.69  & $-$25.96 & 000505  & 0.92 & 0.44  & 58.3 & 16.6 & \\
 & & & & & & & & & & & & \\
145410.1+195648 & 0.243 & 1.9 & 14.17 & 2.23 & 5.33  & $-$26.27 & 000505  & 0.42 & 0.70  & \nodata & \nodata & \\ 
145608.6+275008 & 0.250 & 1 & 13.38 & 2.01 & 2.92  & $-$27.04 & 000505  & 0.17 & 0.31  & \nodata & \nodata  & \\ 
150113.1+232908 & 0.258 & 1.x & 13.46 & 2.48 & 5.84  & $-$27.24 & 000505  & 3.07 & 0.46 & 154.1  & 4.2 & \\
151621.1+225944 & 0.190 & 1.b & 14.12 & 2.11 & 5.28  & $-$25.64 & 000508  & 1.02 & 0.28 & 150.7  & 8.0 & \\
151653.2+190048 & 0.190 & 1 & 11.41 & 2.12 & 4.39  & $-$28.35 & 000508  & 9.37 & 0.08 & 103.5  & 0.3 & spol \\
151901.5+183804 & 0.187 & 1 & 14.25 & 2.15 & 4.65  & $-$25.48 & 000508  & 0.67 & 0.22 & 131.7 & 10.5 & \\
152151.0+225120 & 0.287 & 1.x & 14.30 & 2.33 & 6.20  & $-$26.62 & 000509  & 1.02 & 0.48 & 117.5 & 16.3 & \\
 & & & & & & & & & & & & \\
154307.7+193751 & 0.228 & 1.5 & 12.74 & 2.29 & 3.86  & $-$27.55 & 000505  & 1.33 & 0.26  & 29.6  & 5.6 & \\
163700.2+222114 & 0.211 & 1.x & 13.59 & 2.10 & 5.41  & $-$26.44 & 990913  & 4.91 & 1.89 & 111.0 & 12.4 & 2H60 \\
   & & & & & & & 000505  & 2.27 & 0.41 & 101.2  & 5.1 & \\
   & & & & & & & ave.   &  2.34 & 0.40 & 102.1  & 4.9 & \\
163736.5+254302 & 0.277 & 1.9 & 14.17 & 2.34 & 5.23  & $-$26.66 & 991012  & 5.05 & 3.25 & 109.3 & 25.4 & 2H90 \\
   & & & & & & & 000508  & 0.60 & 0.39  & 53.3 & 25.6 & \\
   & & & & & & & 000509  & 1.34 & 0.41 & 158.5  & 8.7 & \\
   & & & & & & & ave.   &  0.39 & 0.28  & \nodata & \nodata & \\ 
165939.7+183436 & 0.170 & 1.5 & 12.91 & 2.17 & 5.29  & $-$26.59 & 990913  & 6.30 & 0.73 & 162.4  & 3.3 & 2H60;spol \\
170003.0+211823 & 0.596 & 1.5 & 14.88 & 2.46 & 7.21  & $-$28.30 & 000508 & 11.11 & 0.80 & 109.3  & 2.0 & \\
170536.6+210137 & 0.271 & 1.x & 14.31 & 2.48 & 5.99  & $-$26.52 & 991011  & 0.90 & 1.43  & \nodata & \nodata & 2H90 \\
171442.7+260248 & 0.163 & 1 & 13.08 & 2.25 & 3.82  & $-$26.34 & 990911  & 0.86 & 0.33  & 64.5 & 12.1 & 2H60 \\
 & & & & & & & & & & & & \\
171559.7+280717 & 0.524 & 1.8 & 14.63 & 2.46 & $>$6.37  & $-$28.14 & 000509  & 6.08 & 1.28  & 6.8  & 6.0 & \\
222202.2+195231 & 0.366 & 1.5 & 13.30 & 2.88 & 6.20  & $-$28.60 & 991011  & 7.19 & 1.14 & 109.1  & 4.5 & 2H90 \\
222221.1+195947 & 0.211 & 1 & 12.92 & 2.11 & 4.58  & $-$27.10 & 990911  & 0.95 & 0.23 & 138.9  & 7.0 & 2H60 \\
222554.2+195837 & 0.147  & 2 & 13.49 & 2.15 & 5.31  & $-$25.65 & 990911  & 1.38 & 0.56  & 58.6 & 13.4 & 2H60 \\
223742.6+145614 & 0.277 & 1 & 14.00 & 2.05 & 3.40  & $-$26.71 & 990913  & 1.37 & 1.07  & \nodata & \nodata & 2H60 \\
   & & & & & & & 990914  & 2.26 & 1.97  & \nodata & \nodata & 2H60 \\
   & & & & & & & 991012  & 0.65 & 0.87  & \nodata & \nodata & 2H90 \\
   & & & & & & & ave.   &  0.60 & 0.64  & \nodata & \nodata & \\ 
223946.0+192955 & 0.194 & S & 14.67 & 2.12 & 4.23  & $-$25.15 & 991011  & 0.60 & 1.33  & \nodata & \nodata & 2H90 \\
230304.3+162440 & 0.289  & 2/S & 14.67 & 2.35 & 5.73  & $-$26.28 & 991011  & 1.11 & 1.43  & \nodata & \nodata & 2H90 \\
   & & & & & & & 000108  & 0.22 & 0.55  & \nodata & \nodata  & \\ 
   & & & & & & & ave.   &  0.24 & 0.51  & \nodata & \nodata & \\
 & & & & & & & & & & & & \\
230307.2+254503 & 0.331 & 1.5 & 14.50 & 2.09 & 6.21  & $-$26.70 & 991011  & 4.38 & 1.16 & 140.0  & 7.6 & 2H90 \\
230442.4+270616 & 0.237 & 1.5 & 14.77 & 2.13 & 5.03  & $-$25.57 & 991011  & 0.25 & 1.90  & \nodata & \nodata & 2H90 \\
   & & & & & & & 000108  & 0.29 & 0.57  & \nodata & \nodata  & \\ 
   & & & & & & & ave.   &  0.24 & 0.55  & \nodata & \nodata & \\ 
232250.4+261845 & 0.328 & 1 & 14.36  & 2.12 & 3.25  & $-$26.83 & 990913  & 1.68 & 1.83  & \nodata & \nodata & 2H60 \\
   & & & & & & & 991012  & 1.82 & 1.22  & \nodata & \nodata & 2H90 \\
   & & & & & & & ave.   &  1.76 & 1.02 & 135.9 & 25.2 & \\
232745.6+162434 & 0.364 & ? & 14.48 & 2.41 & 5.62  & $-$27.16 & 991011  & 3.43 & 4.73  & \nodata & \nodata & 2H90 \\
   & & & & & & & 991016  & 1.05 & 0.57  & 90.6 & 19.8 & \\
   & & & & & & & ave.   &  1.08 & 0.57  & 90.0 & 18.9 & \\
233200.6+291423 & 0.301 & 1.5 & 14.67 & 2.09 & 3.63  & $-$26.27 & 991012  & 2.29 & 1.46  & 77.4 & 24.7 & 2H90 \\
234259.3+134750 & 0.299 & 1.5 & 14.19 & 2.02 & 4.02  & $-$26.70 & 990911  & 0.51 & 0.46  & \nodata & \nodata & 2H60 \\
234449.5+122143 & 0.199 & 1 & 12.91 & 2.07 & 4.49  & $-$26.95 & 990911  & 1.01 & 0.24 & 127.2  & 6.7 & 2H60 \\
\enddata

\tablenotetext{a}{Redshift and spectral classification are taken from 
\citet{cutri01}.  Classification L = LINER and S = Starburst.
Type~1.x refers to objects with broad H$\alpha\/$ emission lines, but show
no evidence of an H$\beta\/$ emission line.  Type~1.b refers to objects
that show broad H$\alpha\/$, but the spectra do not extend far enough
to the blue
to reach H$\beta\/$.}
\tablenotetext{b}{Data are from the 2MASS Point Source Catalog.}
\tablenotetext{c}{The listed degree of linear polarization has not been
corrected for statistical bias.}
\tablenotetext{d}{2H60: Unfiltered observation obtained with the Two-Holer 
photo-polarimeter using a 4\arcsec\ aperture at the SO Mt. Lemmon 1.5~m
(60-inch) telescope.
2H90: Unfiltered observation obtained with the
Two-Holer instrument using a 2\farcs 9
aperture at the Bok (90-inch) telescope.
spol:  Optical spectropolarimetry of 2MASSI J151653.2+190048 and 2MASSI 
J165939.7+183436 is reported by \citet{smith00a}.
}

\end{deluxetable}


\begin{deluxetable}{lrcrrrrcrrrrc}
\tablecolumns{13}
\tablewidth{0pc}
\tabletypesize{\small}
\tablecaption{Optical Polarimetry of Other AGN Found by 2MASS}
\tablehead{
\colhead{Object}  &
\colhead{$z\/$\tablenotemark{a}} & \colhead{Type\tablenotemark{a}} &
\colhead{$K_s\/$\tablenotemark{b}} & \colhead{$J-K_s\/$\tablenotemark{b}} &
\colhead{$B-K_s\/$\tablenotemark{b}} &
\colhead{$M_K\/$} &
\colhead{UT Date} &
\colhead{$P\/$\tablenotemark{c}} &
\colhead{$\sigma_P$} &
\colhead{$\theta\/$} &
\colhead{$\sigma_{\theta}\/$} &
\colhead{Comments\tablenotemark{d}} \\
\colhead{(2MASSI J)} & \colhead{} & \colhead{} & \colhead{} & \colhead{} &
\colhead{} &
\colhead{} &
\colhead{({\it yymmdd\/})} & \colhead{(\%)} &
\colhead{(\%)} & \colhead{(\dgr )} & \colhead{(\dgr )} & \colhead{}}
\startdata

004117.7+155129 & 0.080 & L & 13.91 & 1.52 & 2.70  & $-$23.70 & 000107  & 0.29 & 0.22 & \nodata & \nodata &  \\
004125.3+134335 & 0.033 & 2/L & 14.32 & 1.50 & 2.88  & $-$21.31 & 000108  & 0.42 & 0.35 & \nodata & \nodata &  \\
004930.9+153216 & 0.240 & 1 & 14.01 & 1.83 & 3.69  & $-$26.24 & 991011  & 0.93 & 0.80  & \nodata  & \nodata & 2H90 \\
005219.9+170241 & 0.250  & 1.b & 14.19 & 1.90 & 3.01  & $-$26.19 & 991012  & 1.28 & 1.08  & \nodata  & \nodata & 2H90 \\
005812.8+160201 & 0.211 & 1.b & 13.30 & 1.92 & 3.60  & $-$26.66 & 990909  & 0.67 & 0.28 &  15.2  & 13.8 & 2H60 \\
024026.6+123715 & 0.321 & 1.5 & 13.90 & 1.91 & 3.80  & $-$27.23 & 990909  & 0.79 & 0.70 & \nodata  & \nodata & 2H60 \\
024140.3+163322 & 0.278 &  2  & 15.06 & 1.79 & 6.64  & $-$25.55 & 991016  & 0.37 & 0.81 & \nodata  & \nodata & 2H60 \\
 & & & & & & & & & & & & \\
030750.1+160931 & 0.325 & 1.8 & 14.84 & 1.72 & 2.96  & $-$26.12 & 990111  & 3.04 &
2.96  & \nodata & \nodata & 2H90 \\
  & & & & & &   & 000107  & 0.64 & 0.44  & \nodata & \nodata& \\
   & & & & & & & ave.   &  0.64 & 0.44  & \nodata & \nodata & \\
031302.2+210714 & 0.094 &  L  & 13.84 & 1.90 & 4.05  & $-$24.18 & 000107  & 1.53 & 0.43 & 147.5  & 8.0 &  \\
092145.7+191812 & 1.800 & 1.b & 14.55 & 2.20 & 6.45  & $-$32.27 & 000108  & 3.45 & 3.19 & \nodata  & \nodata &  \\
095430.4+145250 & 0.173  & 2/L & 14.47 & 1.85 & 4.53  & $-$24.99 & 000108  & 0.78 &
0.42 & 123.8 & 19.6 & \\
135852.5+295413 & 0.113 & 1.b & 12.85 & 1.80 & 4.15  & $-$25.58 & 000505  & 4.83 & 0.24 & 18.5  & 1.4 &  \\
142834.0+213014 & 0.285 & 1 & 14.84 & 1.69 & 2.86  & $-$25.78 & 000505  & 0.22 & 0.36
 & \nodata & \nodata  & \\
 & & & & & & & & & & & & \\
145744.9+202809 & 0.069 &  2/L  & 14.87 & 1.40 & 3.93  & $-$22.39 & 000505  & 0.72 & 0.68 & \nodata  & \nodata &  \\
154443.6+271850 & 0.172  & 2 & 14.67 & 1.59 & 4.63  & $-$24.70 & 000508  & 0.44 & 0.40
  & \nodata & \nodata & \\
155059.3+212808 & 0.373 & 1 & 14.31 & 1.80 & 2.99  & $-$27.04 & 000508  & 0.66 & 0.24
   & 83.3 & 12.7 & \\
170056.0+243928 & 0.509 & 1.5 & 14.26 & 1.71 & 2.54  & $-$27.86 & 990911  & 1.56 &
0.29  & 83.4  & 5.4 & 2H60 \\
225902.5+124646 & 0.199 & 1.x & 14.11 & 1.93 & 4.49  & $-$25.71 & 990913  & 1.90 & 1.18
& 132.2 & 23.7 & 2H60 \\
  &  & & & & &   & 991011  & 0.39 & 1.03  & \nodata & \nodata & 2H90 \\
  & & & & & & & ave.   &  1.03 & 0.78  & \nodata & \nodata & \\
234905.2+183315 & \nodata & AGN?  & 14.05 & 2.37 & 8.09  & \nodata & 991016  & 4.30 & 1.24 & 157.6  & 8.2 &  \\
\enddata

\tablenotetext{a}{Redshift and spectral classification are taken from
\citet{cutri01}.  Classification L = LINER and S = Starburst.
Type~1.x refers to objects with broad H$\alpha\/$ emission lines, but show
no evidence of an H$\beta\/$ emission line.  Type~1.b refers to objects
that show broad H$\alpha\/$, but the spectra do not extend far enough
to the blue (or red)
to include H$\beta\/$.}
\tablenotetext{b}{Data are from the 2MASS Point Source Catalog.}
\tablenotetext{c}{The listed degree of linear polarization has not been
corrected for statistical bias.}
\tablenotetext{d}{2H60: Unfiltered observation obtained with the Two-Holer 
photo-polarimeter using a 4\arcsec\ aperture at the SO Mt. Lemmon 1.5~m
(60-inch) telescope.
2H90: Unfiltered observation obtained with the
Two-Holer instrument using a 2\farcs 9
aperture at the Bok (90-inch) telescope.}

\end{deluxetable}

\clearpage

\oddsidemargin=0.0in
\evensidemargin=0.0in

\begin{deluxetable}{llccrrrr}
\tablecolumns{8}
\tablewidth{0pc}
\tablecaption{Galactic Interstellar Polarization Detected in the Fields of 2MASS AGN}
\tablehead{
\colhead{Field}  & \colhead{Object} &
\colhead{RA (J2000)} & \colhead{Dec (J2000)} &
\colhead{$P\/$\tablenotemark{a}} &
\colhead{$\sigma_P$} &
\colhead{$\theta\/$} &
\colhead{$\sigma_{\theta}\/$}\\ 
\colhead{(2MASSI J)} & & \colhead{($hh$:$mm$:$ss.s\/$)} &
\colhead{($\pm dd$:$mm$:$ss\/$)}
& \colhead{(\%)} &
\colhead{(\%)} & \colhead{(\dgr )} & \colhead{(\dgr )}}
\startdata

004118.7+281640 & GSC 1744-2794 & 00:41:25.6 & +28:15:43 & 0.74 & 0.08 & 117.1 & 3.2 \\
005010.1+280619 & field galaxy & 00:50:10.5 & +28:06:24 & 2.40 & 0.92 & 139.8 & 12.3  \\
005055.7+293328 & GSC 1745-651 & 00:50:45.3 & +29:33:36 & 0.73 & 0.06  & 114.7 & 2.3 \\
023430.6+243835 & GSC 1771-1156 & 02:34:28.3 & +24:37:04 & 1.07 & 0.26  & 114.1 & 6.9 \\
030750.1+160931 & & 03:07:53.7 & +16:10:09 & 0.74 & 0.20 & 143.3 &  7.6 \\
031302.2+210714 & & 03:13:02.6 & +21:07:44 & 1.28 & 0.50 & 115.9  & 12.5 \\
032458.2+174849 & & 03:25:02.8 & +17:49:28 & 0.87 & 0.25 & 117.0  & 8.1  \\
034857.6+125547 & & 03:48:58.6 & +12:55:36 & 1.65 & 0.12 & 53.0 & 2.0 \\
095504.5+170556 & & 09:55:03.5 & +17:05:55 & 1.28 & 0.53 & 84.4  & 13.7 \\
234449.5+122143 & GSC 1173-752 & 23:44:43.0 & +12:20:41 & 0.60 & 0.06 & 111.0 & 2.7 \\
\enddata

\tablenotetext{a}{The listed degree of linear polarization has not been
corrected for statistical bias.}

\end{deluxetable}


\begin{deluxetable}{lcrr}
\tablecolumns{4}
\tablewidth{0pc}
\tablecaption{Correlation of Optical Polarization 
with Color and $M_{K_s}$}
\tablehead{
\colhead{}  & \colhead{Number} &
\colhead{$\tau$\tablenotemark{a}} & \colhead{P$_\tau\/$\tablenotemark{b}} }
\startdata

$P\/$,\ \ \ $J - K_s\/$ & 70 & 0.23 & $4.8 \times 10^{-3}$ \\
$P\/$,\ \ \ $B - K_s\/$ & 70\tablenotemark{c} & 0.30 & $2.5 \times 10^{-4}$ \\
$P\/$,\ \ \ $M_{K_s}\/$ & 70 & $-$0.25 & $2.3 \times 10^{-3}$ \\
\enddata

\tablenotetext{a}{Kendall's $\tau\/$:  A value of 0 indicates no
correlation between parameters.  Values of +1 or $-$1 indicate a
complete correlation or anticorrelation, respectively.}
\tablenotetext{b}{The probability of measuring this value of $\tau\/$
if there is no association between parameters.}
\tablenotetext{c}{A magnitude of $B = 21$ is adopted for four objects
that are not detected on the DSS.}

\end{deluxetable}

\clearpage

\begin{deluxetable}{lccccccc}
\tablecolumns{8}
\tablewidth{0pc}
\tablecaption{General Properties of 2MASS and PG QSO Types}
\tablehead{\colhead{} &
\colhead{Number}  & \colhead{$M_{K_s}\/$\tablenotemark{a}} &
\colhead{$J - K_s\/$\tablenotemark{a}} 
& \colhead{$B - K_s\/$\tablenotemark{a}} &
\colhead{Number with} & NVSS or FIRST & {\sl IRAS\/} \\
\colhead{} & \colhead{} & \colhead{} & \colhead{} & \colhead{} &
\colhead{$P > 3$\%} & detections & detections}
\startdata

2MASS QSOs & 70 & $-$26.6 & 2.1 & 4.8 & 9 & 33 & 15 \\
 & & & & & \\
Type 1 & 22 & $-$26.9 & 2.1 & 3.8 & 2 & 3 & 1 \\
Type 1.5 & 19 & $-$26.7 & 2.1 & 4.1 & 5 & 12 & 6 \\
Type 1.8--2 & 15 & $-$25.8 & 2.2 & 5.6 & 2 & 9 & 1 \\
Type 1.?\tablenotemark{b} & 14 & $-$26.6 & 2.3 & 5.8 & 0 & 9 & 7 \\
 & & & & & \\
\cline{1-8}\\
PG QSOs ($z < 0.6$) & 75 & $-$26.7 & 1.8 & 3.1 & 0 & 38 & 19 \\
 & & & & & \\
Type 1\tablenotemark{c} & 63 & $-$26.7 & 1.8 & 3.1 & 0 & 32 & 13 \\
Type 1.5\tablenotemark{c} & 12 & $-$25.9 & 1.7 & 3.2 & 0 & 6 & 6 \\
\enddata

\tablenotetext{a}{The median value of the sample is listed.}
\tablenotetext{b}{This category includes 13 objects with broad H$\alpha$
emission lines, but either no detected
H$\beta$ emission (10 objects), or the H$\beta$ spectral region was not
covered in the spectra obtained by \citet{cutri01} (3 objects).
One object (2M232745) has an ambiguous classification.}
\tablenotetext{c}{Spectral classification is from \citet{veron01}.}

\end{deluxetable}


\begin{deluxetable}{lcrrrr}
\tablecolumns{6}
\tablewidth{0pc}
\tablecaption{Comparison of Polarization Distributions}
\tablehead{
\colhead{Sample\tablenotemark{a}}  & \colhead{Number} &
$\langle P \rangle$ & $\langle \sigma_P \rangle$ & Range in $P\/$ &
\colhead{P$_{KS}$\tablenotemark{b}} \\
\colhead{} & \colhead{} & \colhead{(\%)} & \colhead{(\%)} &
\colhead{(\%)} &\colhead{}}
\startdata

2MASS & 70 & 1.9 & 0.5 & 0.2--11.1 & $1.0 \times 10^{-8}$ \\
PG    & 75 & 0.6 & 0.2 & 0.1--2.5 & \\
 & & & & & \\
2MASS & 70 & 1.9 & 0.5 & 0.2--11.1 & $1.4 \times 10^{-1}$ \\
BALQSO & 85 & 1.2 & 0.4 & 0.1--7.6 & \\
 & & & & & \\
2MASS & 70 & 1.9 & 0.5 & 0.2--11.1 & $5.7 \times 10^{-2}$ \\
{\sl IRAS\/} HIGs & 19 & 4.3 & 0.4 & 0.3--16.4 & \\
 & & & & & \\
PG    & 75 & 0.6 & 0.2 & 0.1--2.5 & $4.3 \times 10^{-6}$ \\
BALQSO & 85 & 1.2 & 0.4 & 0.1--7.6 & \\
 & & & & & \\
PG    & 75 & 0.6 & 0.2 & 0.1--2.5 & $5.6 \times 10^{-5}$ \\
{\sl IRAS\/} HIGs & 19 & 4.3 & 0.4 & 0.3--16.4 & \\
 & & & & & \\
BALQSO & 85 & 1.2 & 0.4 & 0.1--7.6 & $3.7 \times 10^{-3}$ \\
{\sl IRAS\/} HIGs & 19 & 4.3 & 0.4 & 0.3--16.4 & \\
\enddata

\tablenotetext{a}{Polarization data are from \citet{berriman90} (PG),
\citet{hines94} ({\sl IRAS\/} HIGs),
\citet{schmidt99}, \citet{hutsemekers98}, and \citet{lamy00} (BALQSO).}
\tablenotetext{b}{The probability determined from a KS test that the
two samples are drawn from the
same parent population.}

\end{deluxetable}

\oddsidemargin=0.0in
\evensidemargin=0.0in
\clearpage

\begin{figure}
\vspace{5.7in}
\includegraphics{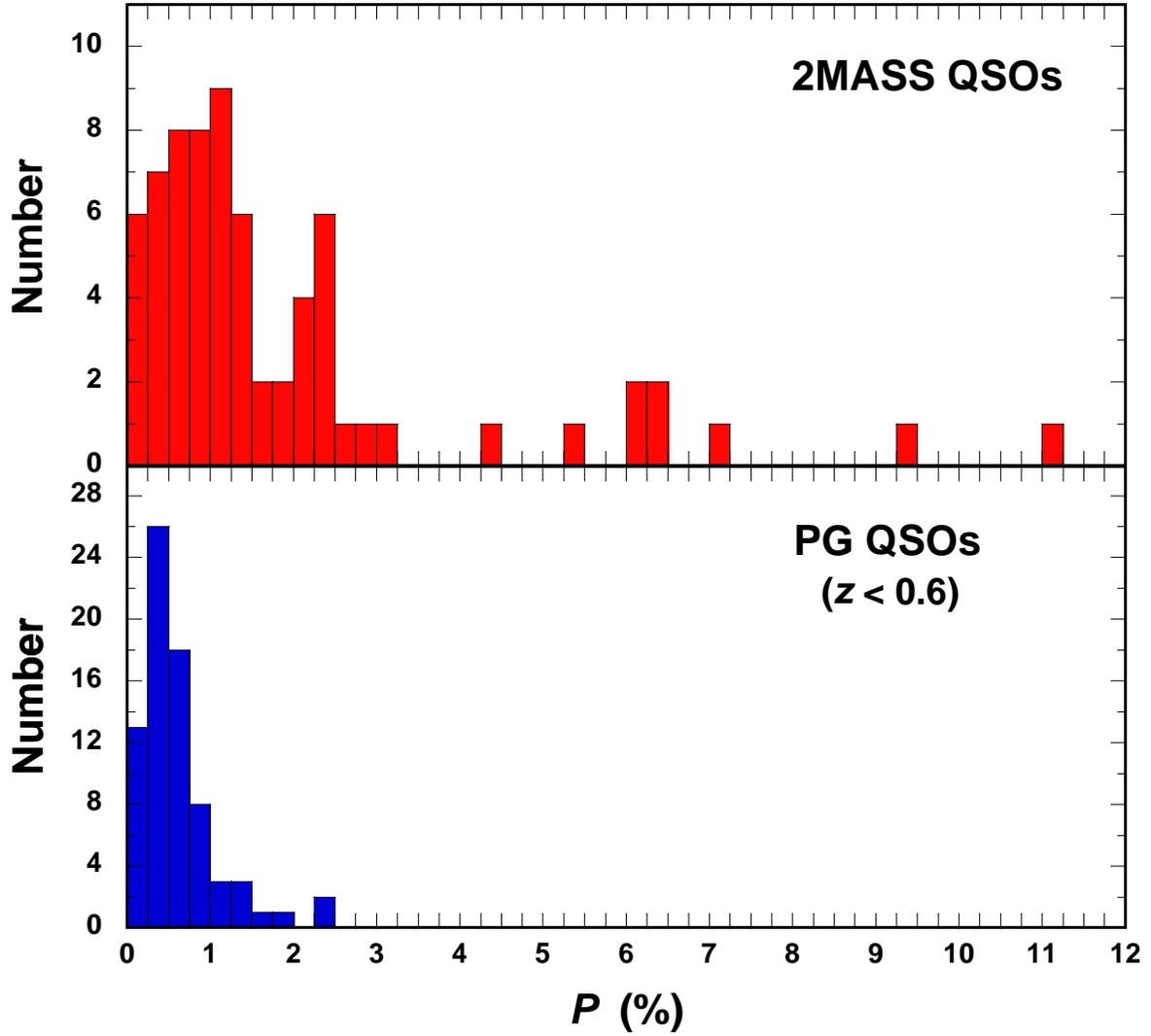}
\caption{The distribution of optical linear polarization
for 2MASS QSOs (top panel) and PG QSOs (bottom panel).
The polarizations have not been corrected for statistical bias and this 
causes the deficiency of objects in the lowest polarization bin for
both samples.
As in all subsequent figures,
only data for PG QSOs with $z < 0.6$ are included, but no object in
this optically selected sample has a polarization above 2.5\%.
\label{fig1}}
\end{figure}
\clearpage

\begin{figure}
\vspace{4.1in}
\includegraphics{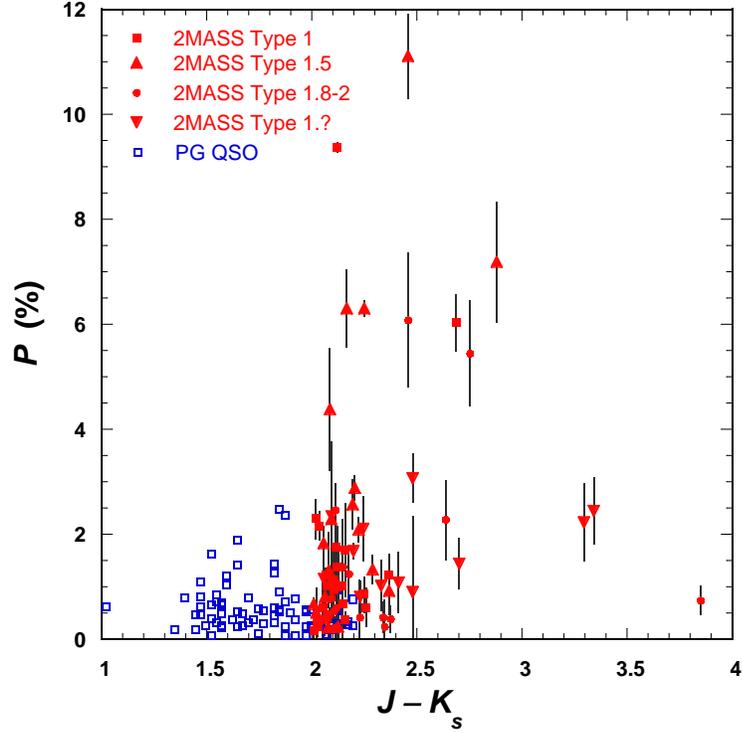}
\caption{Optical polarization (not corrected for statistical
bias) plotted against $J - K_s\/$ color index.
The 2MASS QSOs are denoted by {\it filled\/} symbols and are divided
into four categories according to optical spectral type (see text).
Types~1, 1.5, and 1.8--2 follow the usual convention adopted for Seyfert
galaxies.
``Type~1.?'' include objects of ambiguous type and those
that have broad H$\alpha\/$ emission lines, but
either show no H$\beta\/$ emission, or the available spectra do not
include
H$\beta\/$. 
For comparison, PG QSOs are represented by {\it open\/}
squares.
PG QSO photometric data are from \citet{neugebauer87}.
\label{fig2}}
\end{figure}

\begin{figure}
\vspace{4.1in}
\includegraphics{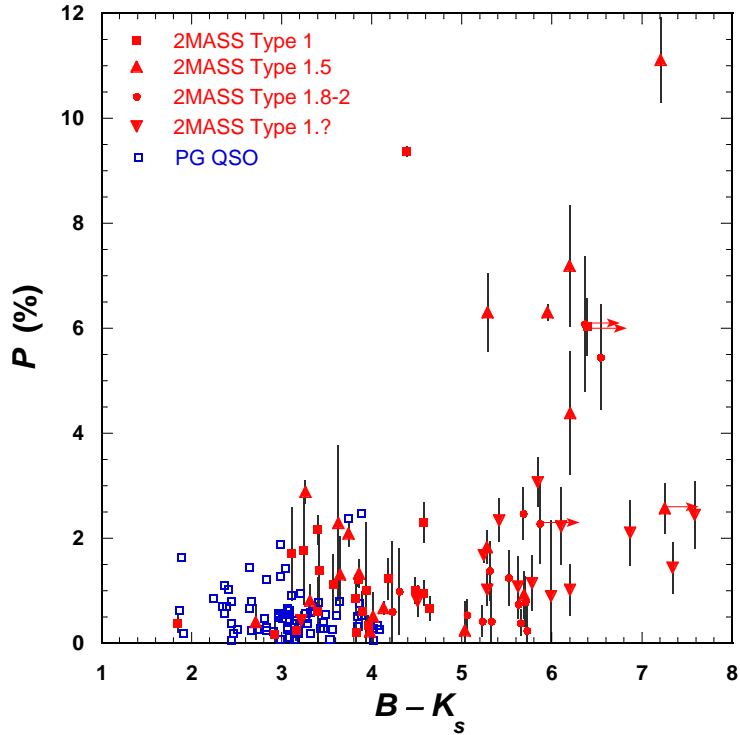}
\caption{Optical polarization plotted against $B - K_s\/$
color index.  
Symbols follow the convention used in Figure~2.
Arrows signify lower limits in $B - K_s\/$ for four 2MASS QSOs. 
\label{fig3}}
\end{figure}
\clearpage

\begin{figure}
\vspace{4.2in}
\includegraphics{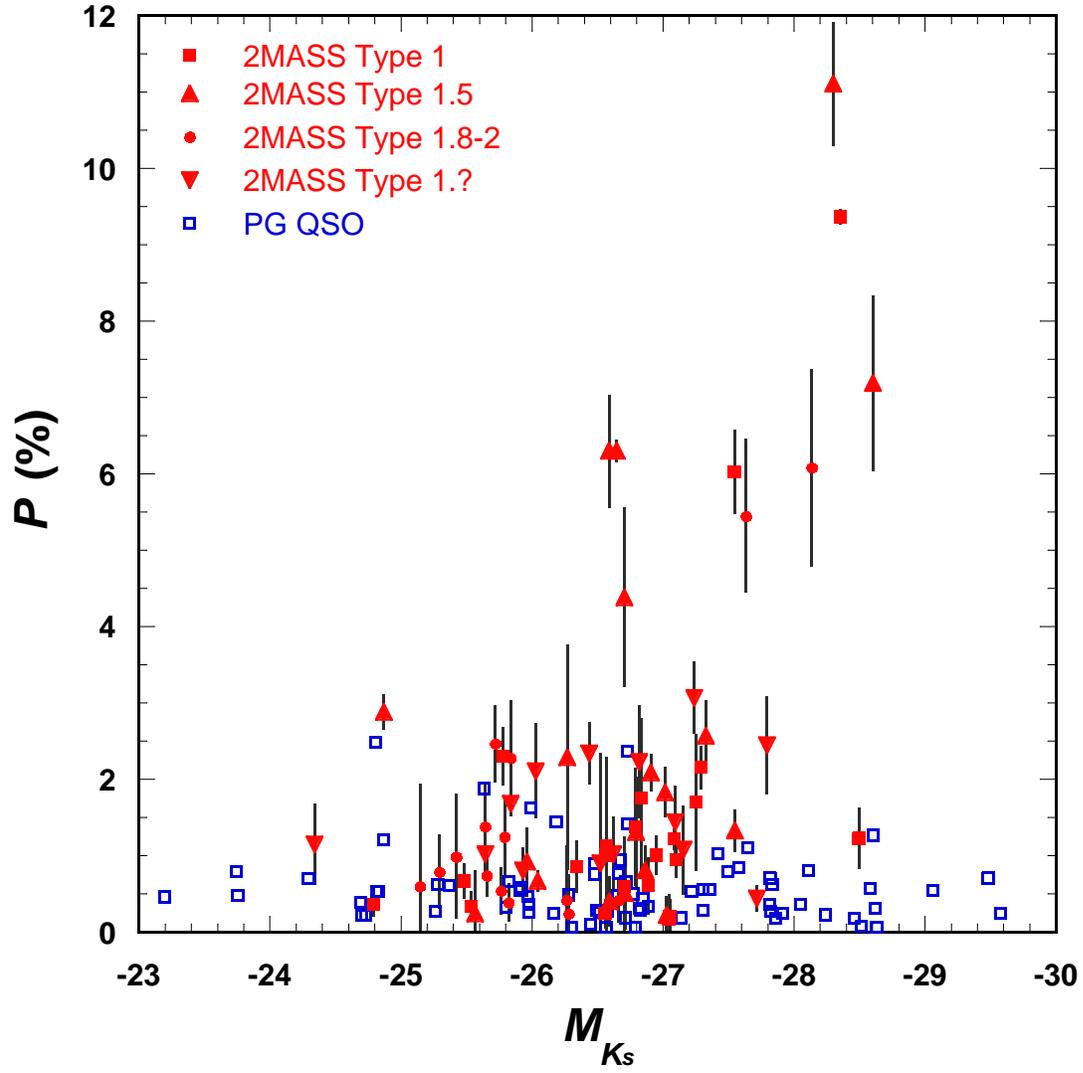}
\caption{Optical polarization plotted against absolute
$K_s\/$ magnitude.
Symbols are the same as in Figure~2 and $M_{K_s}\/$ is K-corrected
using the measured near-IR spectral index.
\label{fig4}}
\end{figure}
\clearpage

\begin{figure}
\vspace{4.1in}
\includegraphics{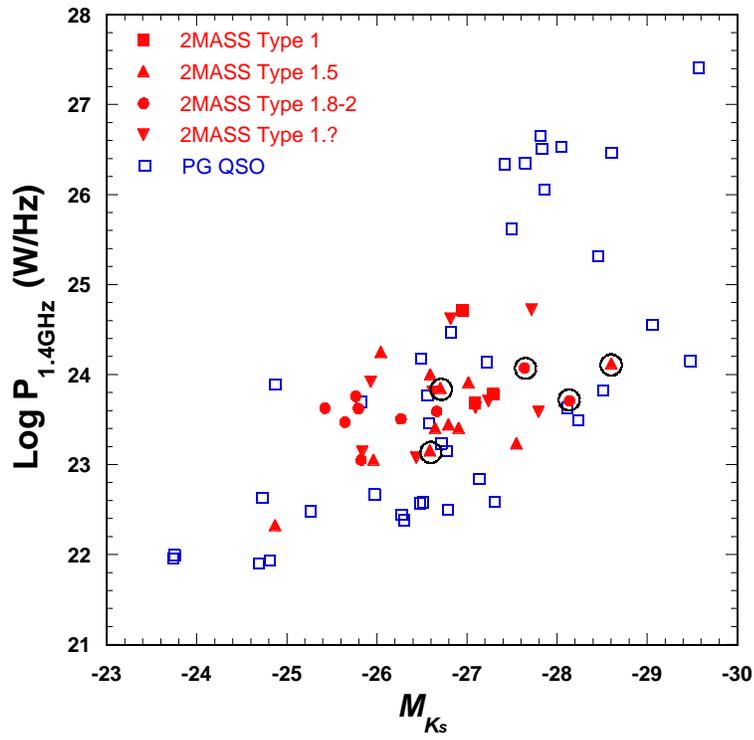}
\caption{Radio power at 1.4~GHz of 2MASS and
PG QSOs detected in either the FIRST or NVSS radio surveys 
plotted against absolute
$K_s\/$ magnitude.
Radio flux densities have been K-corrected assuming a spectral index of 0.
Circled QSOs represent objects with high optical polarization
($P > 3$\%).
\label{fig5}}
\end{figure}

\begin{figure}
\vspace{4.1in}
\includegraphics{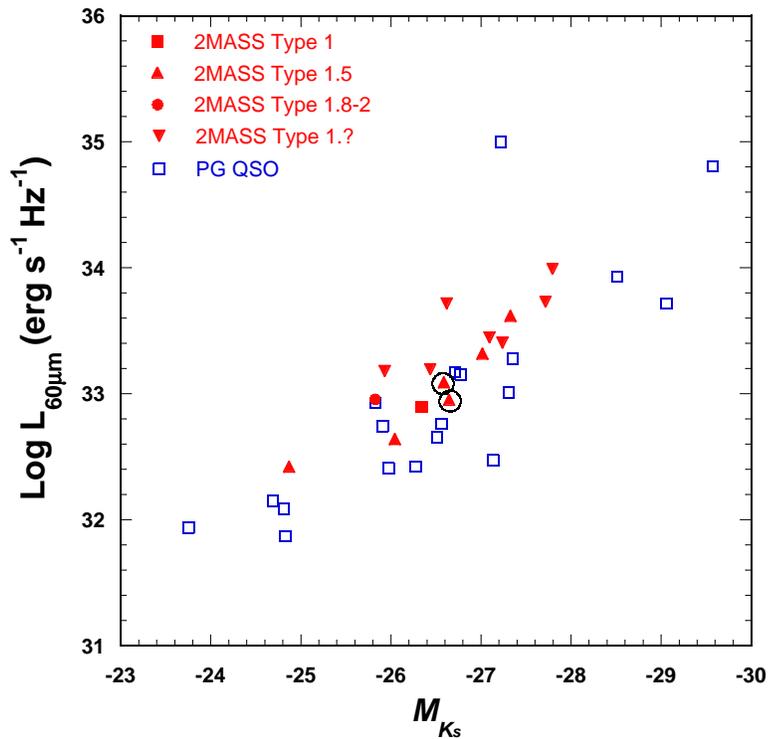}
\caption{The monochromatic 60$\mu$m luminosity of
2MASS and PG QSOs detected by {\sl IRAS\/}.
Circled objects have $P > 3$\%.
\label{fig6}}
\end{figure}
\clearpage

\begin{figure}
\vspace{4.2in}
\includegraphics{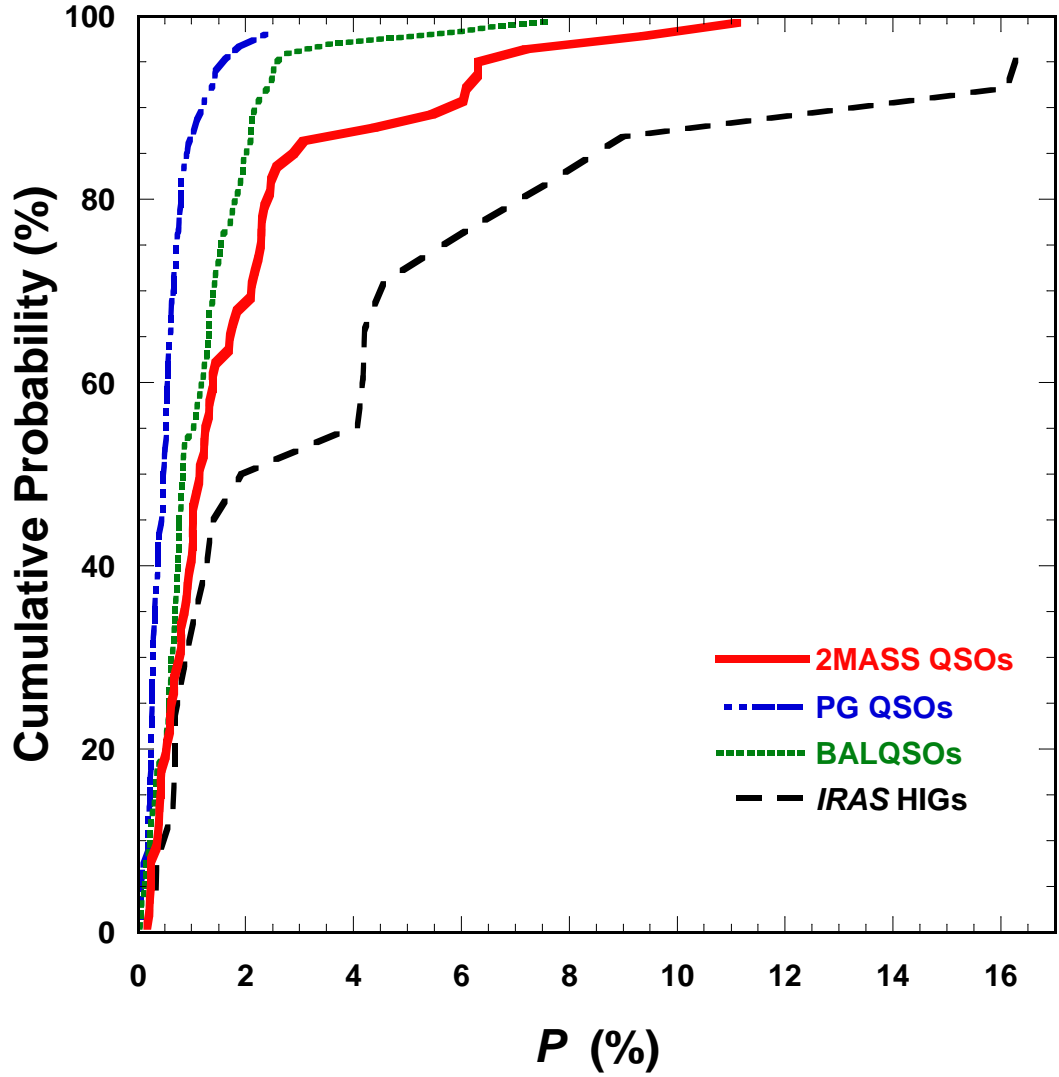}
\caption{The cumulative distributions of optical
polarization for the 2MASS (70 objects),
PG, BALQSO, and {\sl IRAS\/} QSO samples.
Data for the comparison samples are from \citet{berriman90} (PG; 75
objects),
\citet{schmidt99}, \citet{hutsemekers98}, \citet{lamy00} (BALQSO; 85 objects),
and \citet{hines94} ({\sl IRAS\/}; 19 objects).
\label{fig7}}
\end{figure}

\end{document}